# Experimental Investigations on the TMI Thresholds of Low-NA Yb-doped single mode fibers

FRANZ BEIER,[1,2,*] FRIEDRICH MÖLLER,[1] BETTINA SATTLER,[1] JOHANNES NOLD,[1] ANDREAS LIEM,[1] CHRISTIAN HUPEL,[1] STEFAN KUHN,[1] SIGRUN HEIN,[1] NICOLETTA HAARLAMMERT,[1] THOMAS SCHREIBER,[1] RAMONA EBERHARDT[1], AND ANDREAS TÜNNERMANN[1,2]

[1]*Fraunhofer Institute for Applied Optics and Precision Engineering, Albert-Einstein-Str. 7, 07745 Jena, Germany*
[2]*Institute of Applied Physics, Abbe Center of Photonics, Friedrich-Schiller-University Jena, Albert-Einstein-Str. 15, 07745 Jena, Germany*
*Corresponding author: franz.beier@uni-jena.de

**In this contribution we investigate the transversal mode instability behavior of an ytterbium doped commercial 20/400 fiber and obtain 2.9 kW of output power after optimizing the influencing parameters. In this context, we evaluate the influence of the bend diameter and the pump wavelength within the scope of the absorption length and the length of the fiber. Furthermore with a newly developed fiber we report on 4.4 kW of single-mode output power at 40 cm bend diameter.**

Due to their advantages like good beam quality, easy thermal management, high robustness and compact size, fiber lasers are one of the most promising solid state laser concepts for high power scaling with excellent beam quality. Today, single mode power scaling of fiber lasers is still a challenging task. One issue is the reduction of nonlinear effects, especially Raman scattering, which consequently led to increased mode field areas, due to the core intensity and the interaction length scaling thresholds [1]. However, for large mode area fibers, new challenges rise: above a certain threshold, energy is dynamically transferred between different transversal fiber modes resulting in a lack of beam quality and a fluctuating beam profile at the fiber output, namely transversal mode instabilities (TMI) [2]. One of the most influential theories postulates a refractive index grating originating from the interference of amplified transverse modes. As a result, an energy transfer between the transversal modes is possible, which leads to a temporal dynamic of the beam profile. [3]. Over the years, various influencing parameters on the TMI-threshold and/or the grating itself have been identified and investigated theoretically and experimentally [4-11]. While the influence of the bend diameter of the onset threshold of TMI systematically has been investigated before, the impact of the absorption length with regards to the fiber length and the pump wavelength were theoretically, but not systematically investigated in the experiments, yet [5, 11-14].   Furthermore, the influence of the pump direction in terms of bidirectional pumping and fiber coiling was investigated [10, 15, 16]. Both, bidirectional pumping and smaller bend diameters tend to increase the TMI- threshold. The thermally induced change of the refractive index depends on the longitudinal temperature distribution and therefore correlates with the longitudinal power conversion, related to the absorption length. The absorption length itself is determined by the doping concentration, the pump wavelength and the ratio of the fiber core area to the cladding area. For this reason, the dependence of the MI threshold on the fiber length cannot be considered separately from the absorption length or the pump wavelength.

In this contribution we present a systematic investigation of the TMI-threshold in dependence of the bend diameter and the pump wavelength of a well-known commercial fiber with a core diameter of 20µm and a cladding diameter of 400µm. As a result, we increased the output power of the fiber amplifier from below 1 kW up to 3 kW by only tighter bending and a change of the absorption length by a changed pump wavelength. In addition a new transition region is found covering a large power range of 1.3 to 2.8 kW in that fiber. Furthermore, the production of single mode fibers with large core diameters and therefore with extreme low NA is challenging from the technological point of view and reduces the guiding strength of the fiber, which means that the fiber gets very sensitive towards fiber bending [13]. While most power records of single mode fiber amplifier systems are around 3kW, we demonstrated that by careful fiber design, 4.3kW of single mode output power can be obtained [17]. Since the low NA of the fiber under investigation in our previous work led to weak guiding, investigations have only be done for the worst case scenario for TMI of a large bending diameter. Furthermore, it was not completely clarified if mode instabilities could occur at slightly higher power levels. Here, we discuss the stability of 4.4 kW of single-mode output power from a next iteration of the high-power fiber presented in [17] with unprecedented stability, which is bendable down to 40 cm diameter and does not show a transition region compared to the commercial fiber.

**Experimental setup.** The fibers under investigation are employed in a counter pumped amplifier setup as presented in [17]. On both sides of the fiber, a water cooled connector with an anti-reflective coated fused silica endcap was spliced to ensure a stable pump and signal light coupling and for cooling purposes. The fiber itself was cooled by standing water and was coiled to a defined diameter. As seed source, a phase modulated two-tone-single-frequency laser operating at 1067 nm was pre-amplified up to 10 W of output power. The phase-modulation results in a sinusoidal temporal signal, leading to an increased band-width due to self-phase modulation [19]. The fiber was pumped by a diode-laser (Laserline GmbH, wavelength stabilized by water cooling) in counter-propagation direction. The output power beam was separated from the pump light axis by a dichroic mirror. Furthermore, a silica wedge plate was utilized to characterize the output beam and to detect the onset of TMI. The measured output power was corrected with regard to the reflexes at the wedge plate. The TMI-detection was implemented by near field imaging of the output beam on a photo diode (PD). The PD-signal was recorded for 10 s and afterwards split into 500 equal-sized traces. Every single trace was normed to its average value and the standard deviation was calculated. A detailed description of this method was published in Ref. [20].

**Experimental results:** To investigate the influence of the bend diameter on the TMI-threshold, a commercial ytterbium doped fiber (nufern Yb 20/400 Gen.9) with a mode field diameter of 17.7 µm and a numerical aperture of 0.06 was used in a first experiment.

The absorption length of the fiber at 976 nm was theoretically calculated to be 15.7 m for pump absorption of 99 % and 9.7 m for 95 %. The length used in the first experiment was set to the arithmetic mean of 13 m. We gradually reduced the bend diameter from 60 cm down to slightly below 14 cm. Fig. 1 shows the resulting standard deviation and therefore the TMI-thresholds for the bend diameters under investigations.

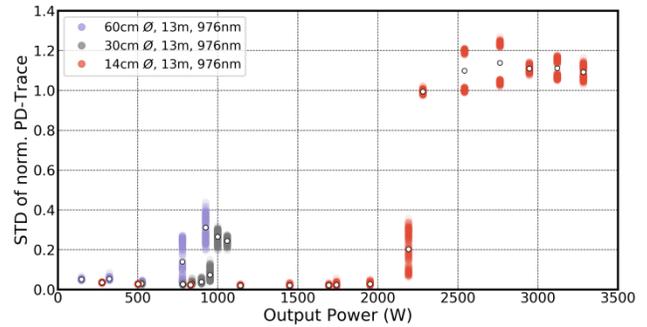

Fig. 1. Standard deviation of normalized PD-Traces for various bend diameters.

As the definition of the TMI-threshold, the five-fold increase of the calculated average standard deviation below the threshold was used as done in Ref. [21]. Applying this definition on the measurement shown in Fig. 1, the TMI threshold is found around 800 W for a bend diameter of 60 cm and is increased up to 1000 W for 30 cm bend diameter. For further bending, the TMI-threshold is rising much stronger, as it is expected for this type of fiber due to the bend dependent modal losses of the higher order modes (HOM) [10, 22]. For 14 cm bend diameter, the TMI-threshold was measured at 2.2 kW in accordance with our previous result [23].

The change of the TMI-threshold within the scope of the fiber length and the pump wavelength was discussed in Ref. [5] and the absorption length of the fiber was stated as a non-negligible variable.

To investigate the influence of the absorption length, we then repeated our experiment with a much longer fiber under similar conditions. From a 30 m fiber, more than twice the fiber length, we obtained a similar TMI-threshold at 976 nm pump wavelength as expected. Fig. 2 shows the standard deviation for both lengths at a pump wavelength of 976 nm in comparison. As a next step, we detuned the pump wavelength from 976 nm to 980 nm in order to increase the pump absorption (99 %) length from 15.7 to 27.0 m. As shown in Fig. 2, we could increase the TMI-threshold to 2.9 kW of signal output power.

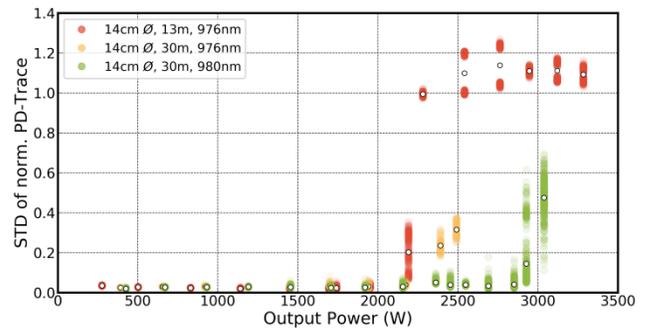

Fig. 2 Standard deviation of normalized PD-Traces for two different fiber and pump wavelengths.

The theoretical and experimental study presented in Ref. [12] investigated the dependence of the TMI-threshold on the pump wavelength as well. There, they chose a fixed amplifier-fiber length that is suitable for all wavelengths under investigation. Their results show that the TMI-threshold is significantly increased by a shift of the pump wavelength from 976nm towards 915nm and 985nm. Unfortunately, the change of the absorption length is only considered in the numerical simulation and compensated by a changed doping concentration. In the experiments, it is not taken into account that a change of the pump wavelength also changes the absorption length of the fiber and thus the differential thermal load. Therefore, the experimental results are not comparable to the numerical simulations they presented. As a result, the pump wavelength is identified as the driving parameter of the TMI-threshold, independent from the thermal load [12]. In contrast, our results show that the TMI threshold was not shifted by a change of the fiber length above the absorption length. Certainly, the pump wavelength significantly changes the pump absorption length of the amplifier fiber system, but also the differential thermal load. Based on the method using a temperature measurement system presented in Ref. [24], we determined the longitudinal thermal load of the fiber amplifier for different output power values. From a rate equation simulation with parameters fitted on our measured thermal load distribution, we moreover determined the maximum thermal load as well as the average thermal load within the absorption length (95% absorbed pump) of the fiber. The thermal load and the absorption lengths in the 3 cases are listed in Table 1 and have to be taken into account.

As can be seen in Table 1, if the coiling and the coupling conditions are maintained, the TMI threshold can be found at a very similar maximum and average thermal load within the absorption length of the fiber. This is a further proof for the significance of the thermal load as a parameter for the prediction of the TMI-threshold within a known fiber. As discussed in [25], a similar average thermal load within the effective length may correlate to a similar grating strength of the predicted thermal grating with regard to the mode overlap of the fundamental mode with HOM, which is maintained in our case. Whereas a comparison of the maximum thermal load compares only a very small length, the average thermal load refers to the total absorption length for 95% pump absorption. The high similarity of the maximum thermal load at the TMI threshold for all three cases could be due to the low longitudinal expansion of the grating strength [3].

**Table 1. Overview of the fiber amplifier parameters for 14 cm bend diameter for two fiber lengths and two pump wavelengths.**

| Fiber Length (m) | pump wavelength (nm) | TMI-Threshold (kW) | Absorption Length (AL95) (m) | Average Thermal Load within AL95 (W/m) | Maximum Thermal Load (W/m) |
|---|---|---|---|---|---|
| 13.0 | 976 | 2.2 | 9.7 | 23.1 | 74.0 |
| 30.0 | 976 | 2.3 | 9.7 | 24.4 | 77.1 |
| 30.0 | 980 | 2.9 | 15.1 | 19.1 | 77.3 |

Furthermore, we identified a semi-stable transition region for few mode step index type fibers below the periodically regime. In this context, the time traces underlying the values shown in green in Fig. 2 were analyzed spectrally over time. Fig. 3 shows the spectrum of the PD-traces and the standard deviation with respect to the output power of the amplifier. In the power range between 1.3 kW and 2.8 kW, the temporal spectrum does not change significantly, while the standard deviation increases with increasing output power. Over 2.8 kW, the periodic regime is reached at a frequency of about 2 kHz. The chaotic range described in Ref. [20] is not reached within the output powers investigated here. Furthermore, we found the described semi-stable transition range in the temporal characteristic of another few-mode fiber with a larger core diameter [24].

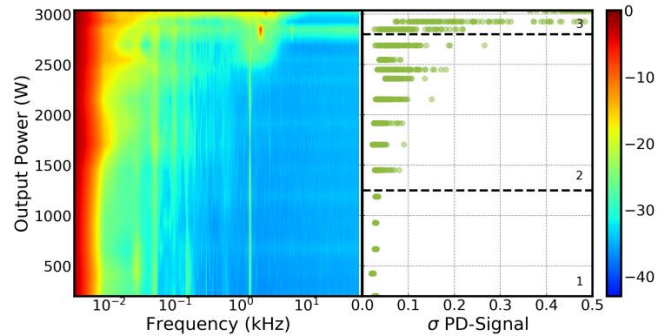

Fig. 3 Spectrogram of the PD-traces for various $P_{out}$ (dB scale). 1, 2, 3 refer to 3 stability regimes: stable, semi-stable and periodic.

In our previous contribution, we showed that we can obtain 4.3 kW of output power from a manufactured fiber with a very low NA [17]. This experiment is again shown in Fig. 4 (blue circles) and show that the standard deviation distribution of a 10 s time trace reaches only a semi-stable regime similar to the nufern-fiber. As a consequence and a further development, we produced a new fiber with a slightly increased NA (0.046 instead of 0.044 and an MFD of 19.4 µm instead of 19.0 µm) to increase the guiding strength.

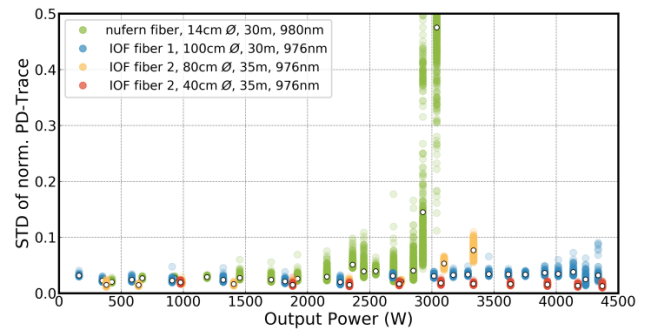

Fig. 4 Standard deviation of the normalized PD-Trace for 3 different fibers.

As the yellow circles show, TMI occur slightly above 3 kW for a bend diameter of 80 cm. Therefore, the bend diameter was reduced to 40cm. In consequence, we reproduced our power record with 4.4 kW, but with a higher stability as the red circle in Fig. 4 show. This result was even achieved for a much smaller bending diameter of 40 cm.

**Summary.** In this paper we have investigated the power scaling potential of the bend diameter and the pump wavelength and their influence on the TMI-threshold of a well-known commercial fiber with a core diameter of 20 µm. We were able to confirm the statement, that the absorption length plays a significant role in comparing the TMI thresholds of different amplifier configurations. The fiber length itself is not a suitable criterion for this, as it is not suitable for determining the distribution of the thermal load, the critical parameter for characterizing the TMI threshold. The TMI-threshold of a 13 m piece of fiber was shifted from 800 W to 2.2 kW by optimizing the bend diameter. At a bend diameter of 14 cm, a fiber length of 30 m and a pump wavelength of 980 nm, we obtained 2.8 kW of stable output power from a commercial fiber. Within the scope of our measurements and a bend diameter of 14 cm, the TMI-threshold correlates with a constant maximum thermal load. Furthermore, we presented the extreme temporal stability of a new iteration of our high power fiber design at a single-mode output power of 4.4 kW.

**Funding.**

BMBF (13N13652), The Fraunhofer and Max Planck cooperation program (PowerQuant), State of Thuringia supported by EU programs EFRE and ESF (2015FOR0017, 13030-715, 2015FGR0107, B715-11011).